\documentclass[aps,prd,twoside,twocolumn,nofootinbib,10pt,superscriptaddress,%
showpacs,floatfix]{revtex4-1}

\usepackage{amsmath,amssymb}
\usepackage{graphicx,bm}
\usepackage{slashed}
\usepackage{dcolumn}
\usepackage{epstopdf}
\usepackage{ulem} 
\usepackage[usenames]{color}
\usepackage{subfigure}
\usepackage[usenames]{color}
\usepackage{tensor}

\allowdisplaybreaks

\begin{document}

\title{\boldmath
$\nuclide[6][\Lambda\Lambda]{He}$ in cluster effective field theory}

\author{Shung-Ichi Ando}%
\email{sando@sunmoon.ac.kr}
\affiliation{Department of Information Display, Sunmoon University, 
Asan, Chungnam 336-708, Korea}

\author{Yongseok Oh}%
\email{yohphy@knu.ac.kr}
\affiliation{Department of Physics, Kyungpook National University, 
Daegu 702-701, Korea}
\affiliation{Asia Pacific Center for Theoretical Physics, Pohang, 
Gyeongbuk 790-784, Korea}

\date{}
%%%%%%%%%%%%%%%%%%%%%%%%%%%%%%%
\begin{abstract}
The hypernucleus $\nuclide[6][\Lambda\Lambda]{He}$ is studied as a three-body 
($\Lambda\Lambda\alpha$) cluster system in cluster effective field theory 
at leading order.  
We find that the three-body contact interaction exhibits the limit cycle when the 
cutoff in the integral equations is sent to the asymptotic limit and thus it 
should be promoted to leading order.
We also derive a determination equation of the limit cycle which reproduces the 
numerically obtained limit cycle.
We then study the correlations between the double $\Lambda$ separation
energy $B_{\Lambda\Lambda}^{}$ of $\nuclide[6][\Lambda\Lambda]{He}$ 
and the scattering length $a_{\Lambda\Lambda}^{}$ of the $S$-wave 
$\Lambda\Lambda$ scattering. 
The role of the scale in this approach is also discussed.
\end{abstract}

\pacs{
21.80.+a, %Hypernuclei
11.10.Hi, %Renormalization group evolution of parameters
21.45.-v  %Few-body systems
}

\maketitle
%%%%%%%%%%%%%%%%%%%%%%%%%%%%%%%%%%%%%%%%%%%%%%%%%%%%%%

Although the first observation of $\nuclide[6][\Lambda\Lambda]{He}$ was reported 
in 1960s~\cite{Prowse66}, there have been only a few reports on this light 
hypernucleus~\cite{DDFM89,TAAA01}.
Among them, a track of $\nuclide[6][\Lambda\Lambda]{He}$ was clearly caught in 
an emulsion experiment of the KEK-E373 Collaboration~\cite{TAAA01},
now known as the ``NAGARA" event, and the two-$\Lambda$ separation energy
$B_{\Lambda\Lambda}$ of $\nuclide[6][\Lambda\Lambda]{He}$ is estimated as 
$B_{\Lambda\Lambda} = 6.93 \pm 0.16$~MeV after being averaged with that from the 
``MIKAGE'' event~\cite{Nakazawa10,E373-13}.
This would be an essential information to study the $\Lambda\Lambda$ interaction.

On the other hand, theoretical studies for double $\Lambda$ hypernuclei mainly aim 
at extracting information on baryon-baryon interactions in the strangeness sector and 
searching for new exotic systems for which the value of $B_{\Lambda\Lambda}^{}$ 
of $\nuclide[6][\Lambda\Lambda]{He}$ plays an important role~\cite{HMRY10,Gal10}. 
Theoretical studies on $\nuclide[6][\Lambda\Lambda]{He}$  have been reported 
with various issues~\cite{th-prl65,AB67,bim-ptp82,cag-npa97,FG02b},
primarily employing the three-body ($\Lambda\Lambda\alpha$) cluster model. 
One of those issues is the role of the mixing of the $\Xi N$ channel 
in the $\Lambda\Lambda$ interaction which is triggered by the small mass difference, 
about 23~MeV, between $\Xi N$ and $\Lambda\Lambda$~\cite{cag-npa97}.

Effective field theories at very low energies are expected to provide a 
model-independent and systematic perturbative method where one introduces a high 
momentum separation scale $\Lambda_H$ between relevant degrees of freedom in 
low energy and irrelevant degrees of freedom in high energy for the system in question.  
Then one constructs an effective Lagrangian expanded in terms of the number of 
derivatives order by order.
Coupling constants appearing in the effective Lagrangian should be determined 
from available experimental or empirical data. 
For a review, see, e.g., Refs.~\cite{BK02,BH04} and references therein.
In the previous publication~\cite{AYO13}, we studied $\nuclide[4][\Lambda\Lambda]{H}$, 
a bound state of a light double $\Lambda$ hypernucleus, and the $S$-wave 
scattering of $\Lambda$ and $\nuclide[3][\Lambda]{H}$ below the hypertriton 
breakup threshold by treating $\nuclide[4][\Lambda\Lambda]{H}$ as a three-body 
($\Lambda$-$\Lambda$-deuteron) system in cluster effective field theory (EFT) at 
leading order (LO).

In this work, we apply this approach to study the structure of 
$\nuclide[6][\Lambda\Lambda]{He}$ as a three-body ($\Lambda\Lambda\alpha$) 
cluster system.
For this purpose, we treat the $\alpha$ particle field as an elementary field.
The binding energy of the $\alpha$ particle is $B_4 \simeq 28.3$~MeV and its 
first excited state has the quantum numbers ($J^\pi = 0^+, I=0$) and the 
excitation energy of $E_1 \simeq 20.0$~MeV, which is between the energy gap of 
$\nuclide[3][]{H}$-$p$ (19.8~MeV) from the ground state energy and that of 
$\nuclide[3][]{He}$-$n$ (20.6~MeV).
Thus the large momentum scale of the $\alpha$-cluster theory is 
$\Lambda_H \simeq \sqrt{2 \mu E_1} \sim 170$~MeV where $\mu$ is the reduced 
mass of the ($\nuclide[3][]{H},p)$ system or the $(\nuclide[3][]{He},n)$ system
so that $\mu \simeq \frac34 m_N^{}$ with $m_N^{}$ being the nucleon mass. 
Therefore, the mixing of the $\Xi N$ channel in the $\Lambda\Lambda$ 
interaction becomes irrelevant because the mass difference $\sim 23$~MeV of the two
channels is larger than $E_1$,  the large energy scale of this approach.
On the other hand, we choose the binding momentum of $\nuclide[5][\Lambda]{He}$
as the typical momentum scale $Q$ of the theory.
The $\Lambda$ separation energy of $\nuclide[5][\Lambda]{He}$ is 
$B_\Lambda \simeq 3.12$~MeV and thus the binding momentum of 
$\nuclide[5][\Lambda]{He}$ as the $(\Lambda\alpha$) cluster system is 
$\gamma_{\Lambda\alpha}^{} = \sqrt{2\mu_{\Lambda\alpha}^{} B_\Lambda}$,
where $\mu_{\Lambda\alpha}^{}$ is the reduced mass of the $\Lambda$-$\alpha$
system.
This leads to $\gamma_{\Lambda\alpha}^{} \simeq 73.2$~MeV and
thus our expansion parameter is
$Q/\Lambda_H \sim \gamma_{\Lambda\alpha}^{}/\Lambda_H \simeq 0.43$.

In addition, a modification of the counting rules is reported by Bedaque, Hammer, 
and van Kolck, which states that the three-body contact interaction should be 
promoted to LO because of the appearance of the ``limit cycle'' in its coupling
in the $S$-wave neutron-deuteron ($nd$) scattering for spin doublet channel in 
the pionless EFT~\cite{BHV99b}.
The limit cycle in a renormalization group analysis was suggested by 
Wilson~\cite{Wilson71} and it is also known that the limit cycle is associated 
with the Efimov states~\cite{Efimov71} in the unitary limit, where the scattering 
length in the $NN$ interaction becomes infinity. 
Furthermore, a ``determination equation'' of the limit cycle, as an expression
of the homogeneous part of the integral equation in the asymptotic limit,
was obtained earlier by Danilov~\cite{Danilov61}.

In this work, we investigate the bound state of the ($\Lambda\Lambda\alpha$) 
cluster system in cluster EFT at LO in order to describe $\nuclide[6][\Lambda\Lambda]{He}$.
We find that the three-body contact interaction exhibits the limit cycle behavior 
when the coupled integral equations with a sharp cutoff are numerically solved. 
Thus the contact interaction should be promoted to LO.
We also derive a determination equation of the limit cycle for the 
$\Lambda\Lambda\alpha$ system and find that the solution of the equation 
reproduces remarkably well the numerically obtained limit cycle.
In addition, we investigate the correlation between $B_{\Lambda\Lambda}$ and 
the scattering length $a_{\Lambda\Lambda}^{}$ of the $S$-wave $\Lambda\Lambda$ 
interaction including the three-body contact interaction with different cutoff values.
The case without the three-body contact interaction will be studied as well.

The LO effective Lagrangian relevant to our study reads
\begin{equation}
\mathcal{L} = \mathcal{L}_\Lambda  + \mathcal{L}_\alpha
+ \mathcal{L}_{s} + \mathcal{L}_{t} + \mathcal{L}_{\Lambda t} .
\end{equation}
Here, $\mathcal{L}_\Lambda$ and $\mathcal{L}_\alpha$ are one-body Lagrangians
for spin-1/2 $\Lambda$ and spin-0 $\alpha$-cluster field in heavy-baryon
formalism~\cite{BKM95,AM98b}, respectively,
\begin{eqnarray}
\mathcal{L}_\Lambda &=& \mathcal{B}_\Lambda^\dagger \left[ i v \cdot D 
+\frac{(v\cdot D)^2 - D^2}{2m_\Lambda^{}} \right] \mathcal{B}_\Lambda 
+ \cdots,
\\
\mathcal{L}_\alpha &=& \phi_\alpha^\dagger
\left[ i v \cdot D +\frac{(v\cdot D)^2 - D^2}{2m_\alpha}\right] \phi_\alpha 
+ \cdots,
\end{eqnarray}
where $v^\mu$ is the velocity vector $v^\mu=(1,\bm{0})$ and $m_\Lambda^{}$ and 
$m_\alpha^{}$ are the $\Lambda$ and $\alpha$ mass, respectively. 
The dots denote higher order terms.
The Lagrangian of the auxiliary fields $s$ and $t$ are given by $\mathcal{L}_s$ 
and $\mathcal{L}_t$, respectively, where $s$ is the dibaryon field of two 
$\Lambda$ particles in $^1S_0$ channel and $t$ is the composite field of the 
$(\Lambda\alpha)$ system in the $\nuclide[5][\Lambda]{He}$ ($S=1/2$) 
channel~\cite{AYO13,BS00,AH04},  
\begin{eqnarray}
\mathcal{L}_{s} &=& 
\sigma_s^{} s^\dagger \left[ i v\cdot \partial 
+ \frac{(v\cdot \partial)^2-\partial^2}{4m_\Lambda^{}}
+ \Delta_s \right] s
\nonumber\\ && \mbox{}
- y_s^{} \left[
s^\dagger \left( \mathcal{B}_\Lambda^T P^{(^1S_0)}\mathcal{B}_\Lambda^{} \right)
+\mbox{H.c.} \right]
+\cdots ,
\\
\mathcal{L}_{t} &=& 
\sigma_t^{} t^\dagger \left[ i v \cdot \partial 
+ \frac{(v\cdot\partial)^2 - \partial^2}{2(m_\Lambda^{} + m_\alpha^{})}
+ \Delta_t \right] t
\nonumber\\ && \mbox{}
- y_t\left[
t^\dagger \mathcal{B}_\Lambda \phi_\alpha + \mbox{H.c.} \right] 
+ \cdots ,
\end{eqnarray}
where $\sigma_{s,t}^{}$ are sign factors.
The mass differences between $\Lambda\Lambda$ and the $s$ dibaryon state
and between $\Lambda\alpha$ and the composite $t$ state ($\nuclide[5][\Lambda]{He}$) 
are represented by $\Delta_{s,t}$, respectively.
$P^{({}^1S_0)} = - i\frac{1}{2}\sigma_2^{}$ is the spin projection operator to the
${}^1S_0$ state.
The dibaryon $s$ state is coupled to two $\Lambda$ in ${}^1S_0$ state 
and the composite $t$ state
to the $S$-wave $\Lambda\alpha$ state with the coupling constants $y_{s,t}^{}$,
respectively.
The Lagrangian for the contact interaction of $\Lambda$ and $t$ reads
\begin{eqnarray}
\mathcal{L}_{\Lambda t} &=& 
- 2 m_\alpha^{} y_t^2 
\frac{g(\Lambda_c)}{\Lambda_c^2} 
\left(\mathcal{B}^T_\Lambda P^{(^1S_0)}t\right)^\dagger
\left(\mathcal{B}^T_\Lambda P^{(^1S_0)}t\right)
+ \cdots ,
\nonumber \\
\end{eqnarray}
where the coupling $g(\Lambda_c)$ is a function of the cutoff $\Lambda_c$,
which is defined in the coupled integral equations below.

In the present work we  consider two composite states 
in the two-body part, namely, the $s$ field and $t$ field.
The dibaryon $s$ state was investigated in our previous publication~\cite{AYO13}, 
where the Feynman diagrams for the dressed dibaryon propagator can be found.
The renormalized dressed dibaryon propagator is obtained as
\begin{equation}
D_{s}(p) = \frac{4\pi}{y_s^2m_\Lambda^{}}
\frac{1}{
\frac{1}{a_{\Lambda\Lambda}^{}}
- \sqrt{-m_\Lambda^{} p_0^{} + \frac14 \bm{p}^2 -i\epsilon }
} ,
\label{eq;Ds}
\end{equation}
where
$y_s = -\frac{2}{m_\Lambda^{}} \sqrt{\frac{2\pi}{r_{\Lambda\Lambda}^{}} }$.
The scattering length and the effective range of $S$-wave $\Lambda\Lambda$ 
scattering are represented by $a_{\Lambda\Lambda}^{}$ and $r_{\Lambda\Lambda}^{}$,
respectively.
We note that the expression of the dressed dibaryon propagator in Eq.~(\ref{eq;Ds}) is
for the large value of $a_{\Lambda\Lambda}^{}$.
In the case of a small value of $a_{\Lambda\Lambda}^{}$ one can expand it in terms
of the kinetic square root term~\cite{vk-npa99}.  
The diagrams for the dressed $t(\Lambda\alpha)$ propagator can be found, e.g., in 
Ref.~\cite{AYO13}, which lead to the renormalized dressed $t(\Lambda\alpha)$ 
propagator as
\begin{eqnarray}
D_t(p) =
\frac{2\pi}{y_t^2 \mu_{\Lambda\alpha}^{}}
\frac{1}{
\gamma_{\Lambda\alpha}^{}
- \sqrt{
-2\mu_{\Lambda\alpha}^{} \left(
p_0^{}
- \frac{1}{2(m_\alpha^{} +m_\Lambda^{})} \bm{p}^2
+ i\epsilon
\right)
}
}\,,
\nonumber \\
\label{eq;Dt}
\end{eqnarray}
where
$y_t^{} = -\frac{1}{\mu_{\Lambda \alpha}^{}}\sqrt{\frac{2\pi}{r_{\Lambda \alpha}^{}}}$.
We also note that the dependence of $y_{s,t}^{}$ on the effective ranges 
$r_{\Lambda\Lambda}^{}$ and $r_{\Lambda\alpha}^{}$ disappears in the final expression 
of the three-body coupled integral equations at LO~\cite{AYO13}.

The amplitude for $S$-wave elastic $\Lambda$-$\nuclide[5][\Lambda]{He}$ scattering 
in CM frame can be described by the coupled integral equations at LO as
\begin{widetext}
\begin{eqnarray}
a(p,k;E) &=& K_{(a)}(p,k;E) 
- m_\alpha^{} y_t^2 \frac{g(\Lambda_c)}{\Lambda_c^2}
- \frac{1}{2\pi^2} \int^{\Lambda_c}_0 dl \, l^2
\left[
K_{(a)}(p,l;E) 
- m_\alpha^{} y_t^2 \frac{g(\Lambda_c)}{\Lambda_c^2}
\right]
D_t \left( E-\frac{l^2}{2m_\Lambda^{}}, \bm{l} \right) 
a(l,k;E)
\nonumber \\ && \mbox{}
- \frac{1}{2\pi^2}\int^{\Lambda_c}_0 dl \, l^2
K_{(b1)}(p,l;E) 
D_s\left( E -\frac{l^2}{2m_\alpha^{}} , \bm{l} \right) 
b(l,k;E) ,
\nonumber \\
b(p,k;E) &=& K_{(b2)}(p,k;E) 
- \frac{1}{2\pi^2}\int^{\Lambda_c}_0 dl \, l^2
K_{(b2)}(p,l;E) 
D_t\left( E -\frac{l^2}{2m_\Lambda} , \bm{l} \right) 
a(l,k;E),
\label{eq;ab}
\end{eqnarray}
\end{widetext}
where the amplitudes $a(p,k;E)$ and $b(p,k;E)$ are half-off shell amplitudes for 
the elastic $\Lambda t$ channel and the inelastic $\Lambda t$ to $\alpha s$ 
channel, respectively.
Here, $p=|\bm{p}|$ and $k=|\bm{k}|$ where $\bm{p}$ ($\bm{k}$) is the off-shell 
final (on-shell initial) relative momentum in the CM frame. 
Thus the total energy $E$ is determined as
$E = \frac{1}{2\mu_{\Lambda(\Lambda\alpha)}^{}} k^2 - B_{\Lambda}$
where $\mu_{\Lambda(\Lambda\alpha)}^{}=m_\Lambda^{}(m_\Lambda^{} + m_\alpha^{})/
(2m_\Lambda^{} + m_\alpha^{})$.
A sharp cutoff $\Lambda_c$ is introduced in the loop integrals of Eq.~(\ref{eq;ab}).
The one-$\alpha$ and one-$\Lambda$ exchange interactions are given by
$K_{(a)}(l,k;E)$ and $K_{(b1,b2)}(l,k;E)$, respectively, where
\begin{eqnarray}
K_{(a)}(p,l;E) &=& 
\frac{m_\alpha^{} y_t^2}{2pl} \ln\left[
\frac{\frac{m_\alpha^{}}{2\mu_{\Lambda\alpha}^{}}(p^2+l^2) + pl - m_\alpha^{} E}
     {\frac{m_\alpha^{}}{2\mu_{\Lambda\alpha}^{}}(p^2+l^2) - pl - m_\alpha^{} E}
\right] ,
\label{eq;Ka}
\nonumber \\
K_{(b1)}(p,l;E) &=& 
\frac{\sqrt2 m_\Lambda^{} y_s^{} y_t^{}}{2pl} \ln\left[
\frac{p^2 + \frac{m_\Lambda^{}}{2\mu_{\Lambda\alpha}^{}}l^2 + pl - m_\Lambda^{} E}
     {p^2 + \frac{m_\Lambda^{}}{2\mu_{\Lambda\alpha}^{}}l^2 - pl - m_\Lambda^{} E}
\right] ,
\label{eq;Kb1}
\nonumber \\
K_{(b2)}(p,l;E) &=&
\frac{\sqrt2 m_\Lambda^{} y_s^{} y_t^{}}{2pl} \ln\left[
\frac{\frac{m_\Lambda^{}}{2\mu_{\Lambda\alpha}^{}}p^2 + l^2 + pl - m_\Lambda^{} E}
     {\frac{m_\Lambda^{}}{2\mu_{\Lambda\alpha}^{}}p^2 + l^2 - pl - m_\Lambda^{} E}
\right] .
\label{eq;Kb2}
\nonumber \\
\end{eqnarray}

%%%%%%%%%%%%% FIG. 1
\begin{figure}[t]
\begin{center}
\includegraphics[width=\columnwidth]{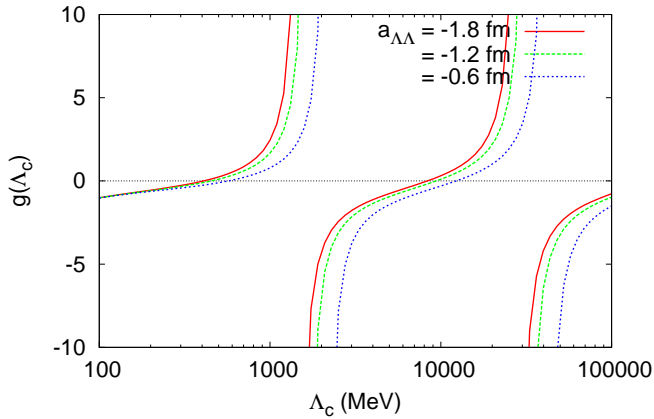}
\caption{(Color Online)
The coupling $g(\Lambda_c)$ as a function of $\Lambda_c$ 
for $a_{\Lambda\Lambda}^{}=-0.6$, $-1.2$, $-1.8$~fm
where the values of $g(\Lambda_c)$ are fitted by 
$B_{\Lambda\Lambda}=6.93$~MeV of $\nuclide[6][\Lambda\Lambda]{He}$. 
}
\label{fig;gvsLam}
\end{center}
\end{figure}
%%%%%%%%%%%%%%%%%%%%%%%%%%%%%%%%%%%%

In Fig.~\ref{fig;gvsLam}, we plot curves of $g(\Lambda_c)$ as a function of 
$\Lambda_c$ with $a_{\Lambda\Lambda}^{} = -0.6$, $-1.2$, $-1.8$~fm.
The curves are numerically obtained from the homogeneous part of the coupled 
integral equations in Eq.~(\ref{eq;ab}) so as to reproduce the three-body bound 
state with $B_{\Lambda\Lambda}=6.93$~MeV. 
One can see that the curves exhibit the limit cycle and the first 
divergence appears at $\Lambda_c \sim 1$~GeV.
In addition, a larger value of $|a_{\Lambda\Lambda}^{}|$ behaves as giving
a larger attractive force and shifts the curves of $g(\Lambda_c)$ to the left in 
Fig.~\ref{fig;gvsLam}.

As pointed out in Ref.~\cite{Griess05}, one can check if the system exhibits the 
limit cycle behavior by studying the homogeneous part of the integral equation 
in the asymptotic limit. 
From Eq.~(\ref{eq;ab}), assuming the form of the amplitude in the asymptotic 
limit $p \gg k$ as $a(p,k) \sim p^{-1-s}$, we have
\begin{equation}
1 = C_1 I_1(s) + C_2 I_2(s) I_3(s) ,
\label{eq:I123}
\end{equation}
where 
\begin{eqnarray}
C_1 = \frac{1}{2\pi}\frac{m_\alpha^{}}{\mu_{\Lambda\alpha}^{}}
\sqrt{
\frac{\mu_{\Lambda(\Lambda\alpha)}^{}}{\mu_{\Lambda\alpha}^{}}},
\quad
C_2 = 
\frac{\sqrt{2 m_\Lambda^{} \mu_{\Lambda(\Lambda\alpha)}^{} 
\mu_{\alpha(\Lambda\Lambda)}^{}
}}{\pi^2 \mu_{\Lambda\alpha}^{3/2}} ,
\nonumber \\
\end{eqnarray}
and $\mu_{\alpha(\Lambda\Lambda)}^{} = 2 m_\Lambda^{} m_\alpha^{} /
(2m_\Lambda^{} + m_\alpha^{})$.  
The functions $I_{1,2,3}(s)$ are obtained by the Mellin transformation~\cite{Ji12}
and their explicit expressions are given in Appendix. 
The imaginary solution $s=\pm i s_0^{}$ indicates the limit cycle solution and we have
\begin{equation}
s_0 = 1.0496\cdots .
\label{eq;s0}
\end{equation}
On the other hand,
the value of $s_0^{}$ can be obtained from the curves of the limit cycle of 
$g(\Lambda_c)$ in Fig.~\ref{fig;gvsLam}. 
The $(n+1)$-th values of $\Lambda_n$ at which $g(\Lambda_c)$ vanishes
can be parameterized as $\Lambda_n = \Lambda_0\exp(n\pi/s_0^{})$. 
By using the second and third vales of $\Lambda_n$ for the three 
values of $a_{\Lambda\Lambda}^{}$, we have 
$s_0^{} = \pi/\ln(\Lambda_2/\Lambda_1)\simeq 1.05$, which is in a very good 
agreement with the value of Eq.~(\ref{eq;s0}).
Furthermore, the value of $s_0^{}$ may be checked by using Fig.~52 in Ref.~\cite{BH04}
which is a plot of $\exp{(\pi/s_0^{})}$ versus $m_1^{} / m_3^{}$ for the mass-imbalanced 
system where $m_1^{} = m_2^{} \neq m_3^{}$. 
In our case, $m_1^{}/m_3^{} = m_\Lambda^{} / m_\alpha^{} \simeq 0.3$, which leads to 
$s_0^{} \simeq 1.05$ by the result of Ref.~\cite{BH04}. 
This is in a very good agreement with what we find in Eq.~(\ref{eq;s0}).

%%%%%%%%%%%%% FIG. 2
\begin{figure}[t]
\begin{center}
\includegraphics[width=\columnwidth]{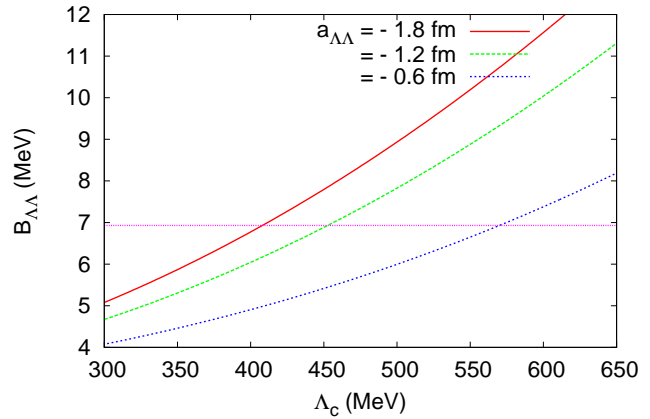}
\caption{(Color Online)
The two-$\Lambda$ separation energy $B_{\Lambda\Lambda}$ as a 
function of the cutoff $\Lambda_c$ for 
$a_{\Lambda\Lambda}^{} = -0.6, -1.2, -1.8$~fm 
without the three-body contact interaction.
The experimental data $B_{\Lambda\Lambda} = 6.93$~MeV is included as a 
reference line.
}
\label{fig;BLL6vsLam-wo-3bf}
\end{center}
\end{figure}
%%%%%%%%%%%%%%%%%%%%%%%%%%%%%%%%%%%%

One may also reproduce the experimental value of $B_{\Lambda\Lambda}$ by 
adjusting the value of $\Lambda_c$ without introducing the three-body contact 
interaction.
In this case, the bound state of $\nuclide[6][\Lambda\Lambda]{He}$ with 
$B_{\Lambda\Lambda}=6.93$~MeV is found to appear only when the cutoff parameter 
$\Lambda_c$ is larger than the critical value $\Lambda_{\rm cr} \approx 300$~MeV,
which is even larger than $\Lambda_H$ of the theory.
We found that $\Lambda_c \approx 300$~MeV leads to 
$a_{\Lambda\Lambda}^{} \approx - 3.4 \times 10^3$~fm.  
When we use $a_{\Lambda\Lambda}^{} =-0.6\sim - 1.8$~fm as obtained 
from the $\nuclide[12]{C}(K^-,K^+\Lambda\Lambda X)$ data in Ref.~\cite{GHH11}, 
we should have $\Lambda_c= 570 \sim 408$~MeV.
In Fig.~\ref{fig;BLL6vsLam-wo-3bf}, we plot $B_{\Lambda\Lambda}$ as a function 
of the cutoff $\Lambda_c$ for $a_{\Lambda\Lambda}^{} = -0.6$, $-1.2$, $-1.8$~fm. 
One can find that $B_{\Lambda\Lambda}$ is quite sensitive to both $\Lambda_c$ 
and $a_{\Lambda\Lambda}$ and it becomes larger as $\Lambda_c$ or 
$|a_{\Lambda\Lambda}|$ increases.

%%%%%%%%%%%%% FIG. 3
\begin{figure}[t]
\begin{center}
\includegraphics[width=\columnwidth]{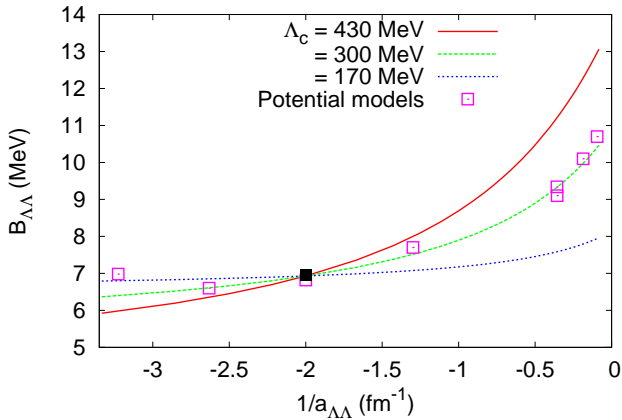}
\caption{(Color Online)
The two-$\Lambda$ separation energy $B_{\Lambda\Lambda}$ as a function 
of $1/a_{\Lambda\Lambda}^{}$ for $\Lambda_c=170$, $300$, $430$~MeV, where 
$g(\Lambda_c)$ is renormalized at the point of $B_{\Lambda\Lambda}=6.93$~MeV 
and $1/a_{\Lambda\Lambda}^{}=-2.0$~fm$^{-1}$ that is marked by a filled square. 
Open squares are the results from the potential models in Table 5 of Ref.~\cite{FG02b}.
}
\label{fig;BLL6vsaLL}
\end{center}
\end{figure}
%%%%%%%%%%%%%%%%%%%%%%%%%%%%%%%%%%%%

Figure~\ref{fig;BLL6vsaLL} shows the two-$\Lambda$ separation energy 
$B_{\Lambda\Lambda}$ as a function of $1/a_{\Lambda\Lambda}^{}$
while $g(\Lambda_c)$ is renormalized at the point marked by a filled square,
i.e., $B_{\Lambda\Lambda}=6.93$~MeV and $1/a_{\Lambda\Lambda}=-2.0$~fm$^{-1}$.
This leads to $g(\Lambda_c) \simeq -0.715$, $-0.447$, $-0.254$ for
$\Lambda_c=170$, $300$, $430$~MeV, respectively.  
Open squares are the estimated values from the potential models given 
in Table 5 of Ref.~\cite{FG02b}.
We find that the curves are sensitive to the cutoff value and the results from 
the potential models are remarkably well reproduced by the curve with 
$\Lambda_c=300$~MeV.

In summary, we have studied the hypernucleus $\nuclide[6][\Lambda\Lambda]{He}$ 
as a three-body $(\Lambda\Lambda\alpha)$ system in cluster EFT at LO.
We found that the three-body contact interaction exhibits the limit-cycle and 
it is needed to be promoted to LO to make the result independent of the cutoff.
The determination equation of the limit cycle for the bound state of 
$\nuclide[6][\Lambda\Lambda]{He}$ is derived and its solutions remarkably well 
reproduce the numerically obtained results for the limit cycle. 
We here note that the determination equation depends on the masses and the 
spin-isospin quantum numbers of the state but not on the details of 
dynamics and that the imaginary solution of the determination equation implies 
the Efimov states in the unitary limit~\cite{BHV99b}. 
Even though the system is not close to the unitary limit, the imaginary solution 
could imply  the presence of a bound state as seen in this study. 
Therefore, the determination equation in three-body cluster systems may be useful 
to search for an exotic state.

We also found that $B_{\Lambda\Lambda}$ of $\nuclide[6][\Lambda\Lambda]{He}$ 
can be reproduced even without introducing the three-body contact interaction, 
which, however, requires $\Lambda_c = 570 \sim 410$~MeV for
$a_{\Lambda\Lambda}^{} = -0.6 \sim -1.8$~fm. 
This range of the cutoff $\Lambda_c$ may be converted to the length scale
$r_c=\Lambda_c^{-1}= 0.35 \sim 0.48$~fm, which overlaps the range of
a hard core potential in the early calculations of Ref.~\cite{AB67}. 
However, the $a_{\Lambda\Lambda}^{}$ dependence is significant and it is 
unlikely to narrow the range of $a_{\Lambda\Lambda}^{}$. 
More precise and diverse experimental data are thus required.

Finally, the correlation between $B_{\Lambda\Lambda}$ and $a_{\Lambda\Lambda}^{}$ 
was investigated by introducing the three-body contact interaction and changing 
the value of $\Lambda_c$.
We find that the results of the potential models can be reasonably reproduced by 
choosing $\Lambda_c =300$~MeV, which may be understood to show the role of 
the two pion exchange as the long range mechanism of the $\Lambda\Lambda$ interaction. 
Meanwhile, choosing $\Lambda_c > \Lambda_H$ ($\Lambda_H\simeq 170$~MeV) is 
inconsistent within our cluster theory because such a large cutoff probes the 
short range (or high momentum) degrees of freedom such as the first excitation 
state of $\alpha$, which is beyond the scope of the present calculation.

\acknowledgments
 
We are grateful to E. Hiyama for suggesting the present work.
We also thank J.K. Ahn for useful discussions and C.~Ji for providing 
his Ph.D. dissertation.
The work of S.-I.A. was supported by the Basic Science Research Program through the 
National Research Foundation of Korea funded by the Ministry of Education under 
Grant No.\ NRF-2012R1A1A2009430.
Y.O. was supported in part by the National Research Foundation of Korea funded 
by the Korean Government (Grant No.\ NRF-2011-220-C00011) and in part
by the Ministry of Science, ICT, and Future Planning (MSIP) and the 
National Research Foundation of Korea under Grant 
No.\ NRF-2013K1A3A7A06056592 (Center for Korean J-PARC Users).

\appendix*

\section{}

The functions $I_{1,2,3}(s)$ in Eq.~(\ref{eq:I123}) are obtained by the Mellin 
transformation~\cite{Ji12} as
\begin{eqnarray}
I_1(s) &=& \int^\infty_0 dx \ln\left( \frac{x^2 +ax +1}{x^2 -ax +1} \right) x^{s-1} 
\nonumber \\ 
&=& \frac{2\pi}{s} \frac{\sin[s\sin^{-1}\left(\frac12a\right)]}{
\cos\left(\frac{\pi}{2}s\right)},
\\
I_2(s) &=& \int^\infty_0 dx \ln\left( \frac{bx^2 +x +1}{bx^2 -x +1} \right) x^{s-1} 
\nonumber \\ 
&=& \frac{2\pi}{s}  \frac{1}{b^{s/2}}
\frac{\sin[s\cot^{-1}\left(\sqrt{4b-1}\right)]}{\cos\left(\frac{\pi}{2}s\right)} ,
\\
I_3(s) &=& \int^\infty_0 dx \ln\left( \frac{x^2 +x +b}{x^2 -x +b} \right) x^{s-1} 
\nonumber \\ 
&=& \frac{2\pi}{s} b^{s/2}
\frac{\sin[s\cot^{-1}\left(\sqrt{4b-1}\right)]}{\cos\left(\frac{\pi}{2}s\right)} ,
\end{eqnarray}
where $a = 2\mu_{\Lambda\alpha}^{}/m_\alpha^{}$ and 
$b=m_\Lambda^{}/(2\mu_{\Lambda\alpha}^{})$.
When $a=b=1$, they reproduce
\begin{eqnarray}
I(s) &=& \int^\infty_0 dx\ln\left(
\frac{x^2 +x +1}
     {x^2 -x +1}
\right) x^{s-1}
= \frac{2\pi}{s} 
\frac{\sin\left(\frac{\pi}{6}s\right)}{
\cos\left(\frac{\pi}{2}s\right)}.
\nonumber \\
\end{eqnarray}


\begin{thebibliography}{10}

\bibitem{Prowse66}
D.~J. Prowse,
\newblock Phys. Rev. Lett. \textbf{17}, 782 (1966).
%%CITATION = PRLTA,17,782;%%

\bibitem{DDFM89}
R.~H. Dalitz, D.~H. Davis, P.~H. Fowler, A.~Montwill, J.~Pniewski, and J.~A.
  Zakrzewski,
\newblock Proc. Roy. Soc. Lond. A \textbf{426}, 1 (1989).
%%CITATION = PRSLA,A426,1;%%

\bibitem{TAAA01}
H.~Takahashi \textit{et~al.\/},
\newblock Phys. Rev. Lett. \textbf{87}, 212502 (2001).
%%CITATION = PRLTA,87,212502;%%

\bibitem{Nakazawa10}
K.~Nakazawa for KEK-E176, E373, and J-PARC E07 Collaborators,
\newblock Nucl. Phys. A \textbf{835}, 207 (2010).
%%CITATION = NUPHA,A835,207;%%

\bibitem{E373-13}
E373 (KEK-PS) Collaboration, J.~K. Ahn \textit{et~al.\/},
\newblock Phys. Rev. C \textbf{88}, 014003 (2013).
%%CITATION = PHRVA,C88,014003;%%

\bibitem{HMRY10}
E.~Hiyama, T.~Motoba, T.~A. Rijken, and Y.~Yamamoto,
\newblock Prog. Theor. Phys. Suppl. \textbf{185}, 1 (2010).
%%CITATION = PTPSA,185,1;%%

\bibitem{Gal10}
A.~Gal,
\newblock Prog. Theor. Phys. Suppl. \textbf{186}, 270 (2010).
%%CITATION = ARXIV:1008.3510;%%

\bibitem{th-prl65}
Y.C. Tang, R.C. Herndon,
Phys. Rev. Lett. \textbf{14}, 991 (1965).

\bibitem{AB67}
S.~Ali and A.~R. Bodmer,
\newblock Nuovo Cim. A \textbf{50}, 511 (1967).
%%CITATION = NUCIA,A50,511;%%

\bibitem{bim-ptp82}
H. Bando, K. Ikeda, and T. Motoba,
Prog. Theor. Phys. \textbf{67}, 508 (1982).
%%CITATION = PTPKA,67,508;%%

\bibitem{cag-npa97}
S.B. Carr, I.R. Afnan, B.F. Gibson,
Nucl. Phys. A \textbf{625} 143 (1997).
%%CITATION = NUPHA,A625,143;%%

\bibitem{FG02b}
I.~Filikhin and A.~Gal,
\newblock Nucl. Phys. A \textbf{707}, 491 (2002).
%%CITATION = NUCL-TH/0203036;%%





\bibitem{BK02}
P.~F. Bedaque and U.~van Kolck,
\newblock Ann. Rev. Nucl. Part. Sci. \textbf{52}, 339 (2002).
%%CITATION = NUCL-TH/0203055;%%

\bibitem{BH04}
E.~Braaten and H.-W. Hammer,
\newblock Phys. Rep. \textbf{428}, 259 (2006).
%%CITATION = COND-MAT/0410417;%%

\bibitem{AYO13}
S.-I. Ando, G.-S. Yang, and Y.~Oh,
\newblock Phys. Rev. C \textbf{89}, 014318 (2014).
%%CITATION = ARXIV:1310.1432;%%

\bibitem{BHV99b}
P.~F. Bedaque, H.~W. Hammer, and U.~van Kolck,
\newblock Phys. Rev. Lett. \textbf{82}, 463 (1999);
%%CITATION = NUCL-TH/9809025;%%
\newblock Nucl. Phys. A \textbf{676}, 357 (2000).
%%CITATION = NUCL-TH/9906032;%%

\bibitem{Wilson71}
K.~G. Wilson,
\newblock Phys. Rev. D \textbf{3}, 1818 (1971).
%%CITATION = PHRVA,D3,1818;%%

\bibitem{Efimov71}
V.~N. Efimov,
\newblock Yad. Fiz. \textbf{12}, 1080 (1970),
\newblock [Sov. J. Nucl. Phys. \textbf{12}, 589--595 (1971)].
%%CITATION = SJNCA,12,589;%%

\bibitem{Danilov61}
G.~S. Danilov,
\newblock Zh. Eksp. Teor. Fiz. \textbf{40}, 498 (1961),
\newblock [Sov. Phys. JETP \textbf{13}, 349--355 (1961)].
%%CITATION = SPHJA,13,349;%%

\bibitem{BKM95}
V.~Bernard, N.~Kaiser, and U.-G. Mei{\ss}ner,
\newblock Int. J. Mod. Phys. E \textbf{4}, 193 (1995).
%%CITATION = HEP-PH/9501384;%%

\bibitem{AM98b}
S.-I. Ando and D.-P. Min,
\newblock Phys. Lett. B \textbf{417}, 177 (1998).
%%CITATION = HEP-PH/9707504;%%

\bibitem{BS00}
S.~R. Beane and M.~J. Savage,
\newblock Nucl. Phys. A \textbf{694}, 511 (2001).
%%CITATION = NUCL-TH/0011067;%%

\bibitem{AH04}
S.-I. Ando and C.~H. Hyun,
\newblock Phys. Rev. C \textbf{72}, 014008 (2005).
%%CITATION = NUCL-TH/0407103;%%

\bibitem{vk-npa99} 
U. van Kolck, Nucl. Phys. A \textbf{645}, 273 (1999).

\bibitem{Griess05}
H.~W. Griesshammer,
\newblock Nucl. Phys. A \textbf{760}, 110 (2005).
%%CITATION = NUCL-TH/0502039;%%

\bibitem{Ji12}
C.~Ji,
\newblock \textit{Universality and beyond: Effective field theory for
  three-body physics in cold atoms and halo nuclei},
\newblock PhD thesis, Ohio University, 2012.

\bibitem{GHH11}
A.~Gasparyan, J.~Haidenbauer, and C.~Hanhart,
\newblock Phys. Rev. C \textbf{85}, 015204 (2012).
%%CITATION = ARXIV:1111.0513;%%


\end{thebibliography}
\end{document}